\documentstyle[aps,epsf,twocolumn]{revtex}
\begin{document}
\twocolumn[\hsize\textwidth\columnwidth\hsize\csname
@twocolumnfalse\endcsname

\title{Coexistence of superconductivity and ferromagnetism in ferromagnetic metals}
\author{N.I. Karchev$^{1,3}$, K.B. Blagoev$^{2,*}$, K.S. Bedell$^3$, and P.B. Littlewood$^2$}
\address{$^1$Department of Theoretical Physics, Faculty of Physics, University of Sofia,
5 James Bourchier Blvd., Sofia 1116, Bulgaria}
\address{$^2$Cavendish Laboratory (TCM), University of Cambridge, Cambridge
CB3 0HE, UK}
\address{$^3$Department of Physics, Boston College, Chestnut Hill, MA
02167}
\date{\today}
\maketitle
\begin{abstract}

In this paper we address the question of coexistence of superconductivity
and ferromagnetism in the high temperature superconductor RuSr$_2$GdCu$_2$O$_{8-\delta}$.
Using a field theoretical approach we study a one-fermion effective model of a 
ferromagnetic superconductor in which the quasiparticles 
responsible for the ferromagnetism
form the Cooper pairs as well. We discuss the physical features which are different
in this model and the standard BCS model and consider their experimental consequences.
\end{abstract}
\pacs{PACS numbers: 71.10.+x, 71.27.+a, 74.10.+v, 75.10.Lp}
]

Recently an itinerant ferromagnet undergoing a high temperature superconducting
transition was discovered in the copper oxide compound RuSr$_2$GdCu$_2$O$_{8-\delta}$ (Ru-1212)
\cite{Bauerfeind et al.1995,C. Bernhard et al.1999,D.J. Pringle et al.1999,Hadjiev et al.1999}
and the experimental studies have revealed that the ferromagnetic state exists even below the 
superconducting transition. This prompts the interesting question of the possible
many-body itinerant fermionic systems supporting both types of broken 
symmetry\cite{Pickett et al.1999}.  

The search for ferromagnetic superconductors goes back to the sixties when 
superconducting materials with magnetic impurities were studied\cite{Abrikosov and Gorkov1960}.
The research in this direction has led to the works of 
Larkin and Ovchinnikov\cite{Larkin and Ovchinnikov64}, and Fulde and 
Ferrell\cite{Fulde and Ferrell64} (LOFF),
who studied a simple model of effective field theory of superconducting
fermions coupled to magnetic impurities and they described the phase diagram of such a 
system.

The Ru-1212 has similarities with the above materials, but is generically different because
the fermions in the RuO$_2$ layers (with mostly ferromagnetic fermions) are hybridized with the
fermions in the CuO$_2$ layers (with mostly superconducting fermions). 
Motivated by the above experimental observations we consider
an effective two band Hamiltonian with ferromagnetic exchange interaction in one
of the bands. One can perform a unitary transformation to diagonalize the kinetic term,
including the tunneling between the two bands. As a result of this hybridization one obtains 
two new fermionic fields. Both fermions have ferromagnetic and pairing interactions (either 
strong/weak or weak/strong) and there are cross terms. The approximation 
of this paper involves neglecting the cross terms. Then one has two 
decoupled one-fermion problems. The one of interest in this paper is for the mostly 
superconducting fermions, because their exchange coupling is much smaller than the exchange 
coupling in the Ru layer, but could be comparable to Tc.

In this paper for the first time the self consistent equations for the superconducting
gap and the magnetization are solved simultaneously in the mean field limit.
We study the one fermion model of a ferromagnetic superconductor leaving 
the two fermion model with the cross terms for further investigation. 
This case is relevant to the doped RuO$_2$ layers in Ru-1212, where the 
magnetism becomes itinerant and the RuO$_2$ layers participate in the 
transport properties of the material. 

Our model Hamiltonian is
\begin{eqnarray}
H-\mu N=\int d^3rc_{\sigma}^{\dagger}(\vec{r})\left(-\frac{1}{2m^*}\vec{\nabla}^2-
\mu\right)c_{\sigma}(\vec{r}) \\
-\frac{J}{2} \int d^3r\vec{S}(\vec{r})\cdot\vec{S}(\vec{r})
-g\int d^3r c_{\uparrow}^{\dagger }(\vec{r})c_{\downarrow}
^{\dagger}(\vec{r})c_{\downarrow}(\vec{r})c_{\uparrow}(\vec{r}), \nonumber
\end{eqnarray}
where $c_{\sigma}(\vec{r})$ are the spin $\sigma$ fermion fields, 
$\vec{S}=\frac12c^{\dagger}_{\sigma}\vec{\tau}_{\sigma\sigma\prime}c_{\sigma\prime}$ is the spin
field, $\tau_i$ are the Pauli matrices, and $\mu$ is the chemical potential. 
The exchange interaction  
is ferromagnetic ($J>0$) and the four fermion interaction is attractive ($g>0$). This is
the simplest model which leads to the coexistence of ferromagnetism and superconductivity.

The partition function of the model can be written as a functional integral over the
Grassmann fields $c(\tau,\vec{r})$ and $\bar c(\tau,\vec{r})$\cite{Negele and Orland1988}. 
We introduce a real vector field $\vec{M}(\tau,\vec{r})$ using
the Hubbard-Stratonovich transformation of the exchange term and a
complex scalar field $f(\tau,\vec{r})$ using a Hubbard-Stratonovich
transformation of the second term in Eq.(1). The vector field describes the
fluctuations of the magnetization, while the complex scalar field the 
superconducting fluctuations.
Performing the Gaussian integral
over the fermionic fields we obtain the partition function of the model as an integral
over $\vec{M}$, $f$ and $\bar f$ which we calculate using the steepest descent around
the mean field solutions $\vec{M}=(0,0,M)$ and $\Delta=g<f>$. Here $M=-<S^z>$ defines the
magnetization of the system. The mean field equations are
\begin{equation}
JM+\frac{\delta F_{eff}}{\delta M}=0,
\qquad\frac{2|\Delta|}{g}+\frac{\delta F_{eff}}{\delta |\Delta|}=0,
\end{equation}
where $F_{eff}$ is the Free energy of a theory with the effective Hamiltonian
\begin{eqnarray}
& & H_{eff}=\sum_{\vec{p}} \left[\epsilon_p^{\uparrow}c_{\vec{p}\uparrow }^{\dagger }c_{\vec{p}\uparrow }+
\epsilon_p^{\downarrow}c_{\vec{p}\downarrow }^{\dagger }c_{\vec{p}\downarrow }+
\bar\Delta c_{-\vec{p}\downarrow}c_{\vec{p}\uparrow}+
\Delta c_{\vec{p}\uparrow }^{\dagger }c_{-\vec{p}\downarrow }^{\dagger }\right] \nonumber \\
& & \epsilon_p^{\uparrow}=\frac{p^2}{2m^*}-\mu+\frac{JM}{2}, 
\quad \epsilon_p^{\downarrow}=\frac{p^2}{2m^*}-\mu-\frac{JM}{2}.
\end{eqnarray}
Here the fermionic effective Hamiltonian $H_{eff}$ is obtained after the Hubbard-Stratonovich 
transformations
and setting the fields at their mean field values.

Next we diagonalize the effective Hamiltonian using a Bogoliubov transformation. After the transformation the
new dispersion relations are
\begin{equation}
E^{\alpha}_p=\frac{JM}{2}+\sqrt{\epsilon^2_p+|\Delta|^2},\,
E^{\beta}_p=\frac{JM}{2}-\sqrt{\epsilon^2_p+|\Delta|^2},
\label{dissp1}
\end{equation}
where $\epsilon_p=\frac{p^2}{2m^*}-\mu$. Then the mean field equations take the form
\begin{equation}
M=\frac12 \int\frac{d^3p}{(2\pi)^3}\left(1-n_p^{\alpha}-n_p^{\beta}\right), \,
\end{equation}
\begin{equation}
1=\frac{g}{2}\int\frac{d^3p}{(2\pi)^3}\frac{n_p^{\beta}-n_p^{\alpha}}{\sqrt{\epsilon^2_p+|\Delta|^2}}
\end{equation}
where $n_p^{\alpha}$ and $n_p^{\beta}$ are the momentum distribution function of the Bogoliubov
fermions and we have assumed that $|\Delta|\neq0$. 

From Eqn.(\ref{dissp1}) one sees that for $M\ge0$ (the convention that we will use here) 
$E_p^{\alpha}>0$ for all momenta $p$ and therefore for $T=0$, $n_p^{\alpha}=0$. For $E_p^{\beta}$
there are two possibilities. When $JM<2|\Delta|$, $E_p^{\beta}<0$ for all $p$ and therefore 
$n_p^{\beta}=1$ for all $p$. Substitution of this in Eqn.(5) leads to $M=0$. Therefore the only
solution of the mean field equations which allows for the coexistence of ferromagnetism and
superconductivity is in the case when $JM>2|\Delta|$ which we will assume. Then the equation
$E_p^{\beta}=0$ has two solutions:
\begin{equation}
p_F^{\pm}=\sqrt{2m^*\mu\pm m^*\sqrt{(JM)^2-4|\Delta|^2}}
\end{equation}
The dispersion of the $\beta$ fermion is positive when $p_F^-<p<p_F^+$ and is negative
in the complementary interval. The $p$ dependence of $E_p^{\beta}$ is depicted in Fig.1. 
With this in mind the Eq.(5) and (6) at $T=0$ have the form
\begin{equation}
M=\frac{1}{12\pi^2}\left[\left(p_F^+\right)^3-\left(p_F^-\right)^3\right],
\end{equation}
\begin{equation}
1=\frac{g}{(2\pi)^2}\left(\int_{0}^{\infty}dp\frac{p^2}{\sqrt{\epsilon^2_p+|\Delta|^2}}
-\int_{p_F^-}^{p_F^+}dp\frac{p^2}{\sqrt{\epsilon^2_p+|\Delta|^2}}\right).
\end{equation}

It is difficult to solve analytically these equations, however
when $JM$ is greater, but close to $2\Delta$, $p_F^+$ is approximately equal to $p_F^-$ and
therefore $M$ is small as follows from Eqn.(8). In this case one can expand the r.h.s.
of Eqn.(7) in the small parameter $\sqrt{(JM)^2-4|\Delta|^2}$ obtaining
\begin{equation}
p_F^{\pm}=p_F\pm\frac{m^*}{2p_F}\sqrt{(JM)^2-4|\Delta|^2},
\end{equation}
where $p_F=\sqrt{2\mu m^*}$. Substitution of these expressions in Eqn.(8)
shows that in this approximation the magnetization is linear in $|\Delta|$, namely
\begin{equation}
M=\frac{2}{J}\frac{r}{\sqrt{r^2-1}}|\Delta|,
\end{equation}
where $r=Jm^* p_F/4\pi^2$ and this expression is valid for large $r$ 
(i.e. $MJ-2|\Delta|\rightarrow 0^+$).

As in the standard BCS theory of superconductivity, the pairing of the quasiparticles
occurs in the vicinity of $p_F$, which must include the interval between $p_F^-$ and
$p_F^+$. Then the  integration in the first integral 
on the r.h.s. of Eqn.(9) is limited to a shell of width $2\Lambda$, i.e.
\begin{equation}
1=\frac{g}{(2\pi)^2}\left(\int_{p_F-\Lambda}^{p_F+\Lambda}dp
\frac{p^2}{\sqrt{\epsilon^2_p+|\Delta|^2}}
-\int_{p_F^-}^{p_F^+}dp\frac{p^2}{\sqrt{\epsilon^2_p+|\Delta|^2}}\right).
\end{equation}
Here $p_F+\Lambda>p_F^+$ and $p_F-\Lambda<p_F^-$.

\begin{figure}[h]
\vspace{0.3cm}
\epsfxsize=7.0cm
\hspace*{0.2cm}
\epsfbox{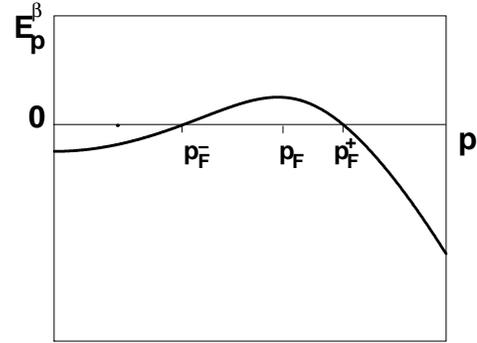}
\caption{$E_p^{\beta}$ as a function of $p$.}
\label{fig1}
\end{figure}

Substitution of the approximate expressions for $p_F^{\pm}$ from Eqn.(10) in
the second term in the r.h.s. of Eqn.(12) and 
using the expression for the magnetization from Eqn.(11) we see that  this term is 
independent of $|\Delta|$ and only leads to
a renormalization of the cutoff $\Lambda$. In that approximation the solution is
\begin{equation}
|\Delta|=\sqrt{\frac{r-1}{r+1}}\Lambda e^{-\frac{2\pi^2}{gm^*p_F}},
\end{equation}
\begin{equation}
M=\frac{2}{J}\frac{r}{r+1}\Lambda e^{-\frac{2\pi^2}{gm^*p_F}}.
\end{equation}
When the magnetization increases the domain of integration in the second
integral on the r.h.s. of Eqn.(12) can exceed the size of the domain around
$p_F$ in which the pairing occurs and which is the integration domain in the
first integral of the same equation. In that case the second integral
dominates and this leads to the absence of solutions with a finite gap.
Taking the limiting case when the two integration domains are equal, i.e.
$p_F+\Lambda=p_F^+$ and $p_F-\Lambda=p_F^-$ where $p_F^\pm$ are the values
of the momenta from Eqn.(10) with $\Delta=0$, we obtain the critical
value of the magnetization 
\begin{equation}
M_c=\frac{\Lambda}{m^*J}(2p_F+\Lambda)
\end{equation}
above which the superconductivity disappears besides the existence of an attractive
four fermion interaction.

Next we calculate the distribution functions $n_p^{\uparrow}$ and $n_p^{\downarrow}$ 
of the spin up and spin down quasiparticles. In terms of the distribution
functions of the Bogoliubov fermions these momentum distribution functions are
\begin{eqnarray}
n_p^{\uparrow}=u_p^2 n_p^{\alpha}+v_p^2n_p^{\beta}, \nonumber \\
n_p^{\downarrow}=u_p^2(1-n_p^{\beta})+v_p^2(1-n_p^{\alpha}),
\end{eqnarray}
where $u_p^2$ and $v_p^2$ are the coefficients in the Bogoliubov transformation. They
are independent of the magnetization and have the same form as in the BCS theory.

At zero temperature $n_p^{\alpha}$ is zero and 
$n_p^{\beta}=\theta(p_F^--p)+\theta(p-p_F^+)$. Then the spin up and spin down 
quasiparticles have the following momentum distribution functions

\begin{equation}
n_p^{\uparrow}=v_p^2\left[\theta(p_F^--p)+\theta(p-p_F^+)\right],
\end{equation}
\begin{equation}
n_p^{\downarrow}=\theta(p_F^+-p)-\theta(p_F^--p)+
v_p^2\left[\theta(p_F^--p)+\theta(p-p_F^+)\right].
\end{equation}
The functions are depicted on Fig.2. 

The appearance of the Fermi surfaces of the Bogoliubov fermion $\beta$ is unexpected in
the superconducting phase, but it is a necessary condition for the existence of itinerant
ferromagnetism. Therefore in the case of coexistence of superconductivity and ferromagnetism
caused by the same quasiparticles the existence of the two Fermi surfaces is a generic
property of this state. These Fermi surfaces are reflected in the spin up and spin down 
momentum distribution functions as well as in the anomalous Green's functions. It is easy
to show that the anomalous Green's function,
\begin{equation}
{\cal F}(\tau - \tau',\vec{p})=-\left<Tc_{\downarrow}
(\tau,-\vec{p})c_{\uparrow}(\tau',\vec{p})\right>,
\end{equation}
in the case $\tau=\tau'$ is
\begin{equation}
{\cal F}(0,\vec{p})=\frac{|\Delta|}{2\sqrt{\epsilon_p^2+|\Delta|^2}}
\end{equation}
when $0<p<p_F^-$ and $p>p_F^+$ and is zero when the momentum $p$ is between the two Fermi surfaces
$p_F^-$ and $p>p_F^+$.

\begin{figure}[h]
\vspace{0.3cm}
\epsfxsize=7.0cm
\hspace*{0.2cm}
\epsfbox{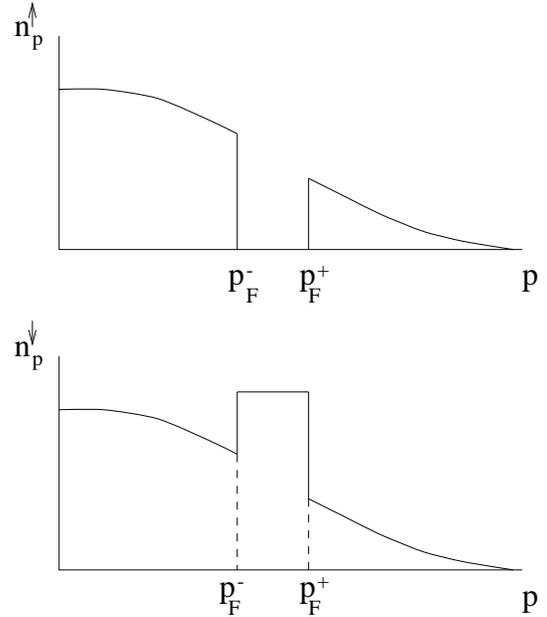}
\caption{The zero temperature momentum distribution functions for spin up and spin down
fermion.}
\label{fig}
\end{figure}

The existence of the Fermi surfaces, leads to different thermodynamic
properties of the system, compared to the standard BCS theory. 
The specific heat has a linear temperature dependence at low temperatures
as opposed to the exponential decrease of the specific heat in the BCS theory
\begin{equation}
C=\frac{2\pi^2}{3}N(0)T.
\end{equation}
Here
\begin{equation}
N(0)=\frac{m^*}{4\pi^2}\frac{p_F^+ + p_F^-}{\sqrt{1-\frac{4|\Delta|^2}{J^2M^2}}}=N^+(0)+N^-(0)
\end{equation}
is the sum of the density of states on the two Fermi surfaces of the Bogoliubov fermion $\beta$.
When the magnetization is small, from Eqns.(10) and (11) follow 
that the density of states increases with $r$ as
$N(0)\rightarrow\frac{m^*}{2\pi^2}rp_F$, as opposed to the case of ordinary weak ferromagnets,
where the density of states is $N(0)=\frac{m^*}{2\pi^2}p_F$ in this limit. 
Hence, the specific heat is large even at very low temperatures.
In the case of a superconductor in an external magnetic field
\cite{Abrikosov and Gorkov1960}
although there are gapless fermionic excitations the specific heat
is not linear as opposed to our case.
This can also be contrasted with some of the unconventional superconductors
which have power law dependence of the specific heat on the temperature, depending
on the nodal structure of the gap function.

Another consequence of the existence of the Fermi surfaces is the existence of
paramagnons which describes the longitudinal spin fluctuations\cite{Izuyama et al.1963}.
They exist in ferromagnetic normal metals and in our theory they survive even in the
ferromagnetic superconducting phase. Their propagator is given by
\begin{equation}
D_l(\omega,p)=\frac{1}{\delta+a\frac{|\omega|}{p}+bp^2},
\end{equation}
where $a$, $b$, and $\delta$ are constants.
The constant 
\begin{equation}
a=\frac{J\pi}{4}\left(1-\frac{4|\Delta|^2}
{J^2M^2}\right)^{-1/2}\left(\frac{N^+(0)}{v_F^+}+\frac{N^-(0)}{v_F^-}\right)
\end{equation}
defines the analytical properties of the paramagnon and is different
from zero because of the existence of the Fermi surfaces. The constant $\delta$ is 
\begin{equation}
\delta=1-\frac{J}{2}N(0)
\end{equation}
and $b$ is a positive constant. As we mentioned earlier, the density of states,
Eqn.(22), increases as the magnetization, $M$ decreases and therefore, at small,
but finite value of the magnetization, $M=M_0$, $\delta$ becomes zero, as opposed to the
weak ferromagnetic metals where $\delta$ becomes zero at zero magnetization.
In the case of coexistence of the superconductivity and ferromagnetism the
superconductivity prevents the magnetization from becoming arbitrarilly small, because
when the magnetization is smaller than the critical value $M_0$, $\delta$
is negative and the paramagnon fluctuations lead to an instability of that phase.
The superconducting phase, with zero magnetization (BCS like regime) the
spin fluctuations of the paramagnon type are absent.    

Recently, a band structure calculation was performed by 
Pickett et al.\cite{Pickett et al.1999} 
and they have studied the origin of the superconducting state in the Ru-1212 compound. 
In our paper we considered the possibility of the coexistence of ferromagnetism and
superconductivity and the physical features of such a system.
 We arrived at a system of self consistent equations for the
magnetization and the superconducting gap, and solved analytically these equations
at small magnetizations. The solutions with coexistence of superconductivity and
ferromagnetism describe Bogoliubov fermions one of which has two Fermi surfaces. 
Therefore the spin up and spin down quasiparticles have two Fermi surfaces each. The
thermodynamic properties of the coexistence phase are different from the standard
BCS theory. The specific heat has a linear temperature dependence as in normal
ferromagnetic metals, but increases anomalously at small magnetizations. These
results are obtained in a mean field approximation, but they are generic for the
coexistence state and can be used as a starting point for a calculations
beyond mean field.    
In our model the quantum critical point is dressed, i.e. the superconducting
state occurs at zero magnetization, because the superconducting
gap is generated not by the spin fluctuations, but by some other means. This
is to be contrasted with the theory of spin fluctuations mediated pairing
in weak ferromagnetic metals 
\cite{Blagoev et al.1999} where the quantum critical point is naked and the
superconducting ferromagnetic critical temperatures go to zero at the
quantum critical point. In this paper we have considered only uniform states.
However, periodic solutions (like the LOFF state in the magnetic impurity
case) will likely exist for certain regions of parameters, but possibly
involving periodic magnetic structures as well as modulated superconducting
order parameter.

NIK and KSB was sponsored by the US National Science Foundation Grant INT9876873.
KBB and PBL were sponsored by the EPSRC Grant GR/L55346.
KSB was partly sponsored by the DOE Grant DEFG0297ER45636.

\end{document}